\documentclass[10pt, conference, compsocconf]{IEEEtran}
\usepackage{cite}

\ifCLASSINFOpdf
   \usepackage[pdftex]{graphicx}
\else
\fi
\usepackage{url}

\usepackage{framed}
\usepackage{colortbl}
\definecolor{mygray}{gray}{0.7}
\definecolor{darkblue}{cmyk}{0.7451, 0.2745, 0.1373,0.0392}
\definecolor{Gray}{gray}{0.85}

\usepackage{bibspacing}
\begin{document}
%
\title{The Quamoco Product Quality Modelling and Assessment Approach}

\author{\IEEEauthorblockN{Stefan Wagner\IEEEauthorrefmark{1},
Klaus Lochmann\IEEEauthorrefmark{2}, 
Lars Heinemann\IEEEauthorrefmark{2},
Michael Kl\"as\IEEEauthorrefmark{3}, 
Adam Trendowicz\IEEEauthorrefmark{3},
Reinhold Pl\"osch\IEEEauthorrefmark{4},\\
Andreas Seidl\IEEEauthorrefmark{5},
Andreas Goeb\IEEEauthorrefmark{6}, and
Jonathan Streit\IEEEauthorrefmark{7}
}

\IEEEauthorblockA{\IEEEauthorrefmark{1}Inst. of Software Technology,
University of Stuttgart,
Stuttgart, Germany,
stefan.wagner@informatik.uni-stuttgart.de}

\IEEEauthorblockA{\IEEEauthorrefmark{2}Institut f\"ur Informatik,
Technische Universit\"at M\"unchen,
Garching, Germany,
lochmann,heineman@in.tum.de}

\IEEEauthorblockA{\IEEEauthorrefmark{3}Fraunhofer IESE,
Kaiserslautern, Germany,
michael.klaes,adam.trendowicz@iese.fraunhofer.de}

\IEEEauthorblockA{\IEEEauthorrefmark{4} Department of Business Informatics,
Johannes Kepler University Linz,
Linz, Austria,
reinhold.ploesch@jku.at}

\IEEEauthorblockA{\IEEEauthorrefmark{5} CSD Research,
Capgemini,
Munich, Germany,
andreas.seidl@capgemini.com}

\IEEEauthorblockA{\IEEEauthorrefmark{6} SAP Research,
SAP AG,
Darmstadt, Germany,
andreas.goeb@sap.com}

\IEEEauthorblockA{\IEEEauthorrefmark{7} itestra GmbH,
Munich, Germany,
streit@itestra.com}
}

\maketitle

\begin{abstract}

Published software quality models either provide abstract quality attributes or concrete quality 
assessments. There are no models that seamlessly integrate both aspects. 
In the project Quamoco, we built a comprehensive approach with the aim to close
this gap. 

For this, we developed in several iterations a meta quality model specifying general concepts,
a quality base model covering the most
important quality factors and a quality assessment approach. 
The meta model introduces the new concept of a product factor, which bridges the gap between 
concrete measurements and abstract quality aspects. Product factors have measures and 
instruments to operationalise quality by measurements from manual inspection and tool 
analysis. The base model uses the ISO 25010 quality attributes, which we refine by 200 factors 
and 600 measures for Java and C\# systems.

We found in several empirical validations that the assessment results fit to the expectations of
experts for the corresponding systems. The empirical analyses also showed that several of 
the correlations are statistically significant and that the maintainability part of the base model has the
highest correlation, which fits to the fact that this part is the most comprehensive.
Although we still see room for extending and improving the base model, it shows
a high correspondence with expert opinions and hence is able to form the basis for 
repeatable and understandable quality assessments in practice.
\end{abstract}


\IEEEpeerreviewmaketitle

\section{Introduction}

Despite great efforts in both research and practice, software quality
continues to be a controversial and insufficiently understood issue
and the quality of software products is often unsatisfactory.
Economical impacts are enormous and include
not only spectacular failures of software but also increased
maintenance costs, resource consumption, long test cycles and user waiting times.

\subsection{Quality Models -- Benefits and Shortcomings}
Software quality models (QMs) tackle these issues by providing a systematic approach for
modelling quality requirements, analysing and monitoring quality
and directing quality improvement measures \cite{2009_deissenboeckf_purposes_scenarios}. 
They thus allow to ensure quality early in the development process. 

In practice, however, a gap remains between two different types of QMs: Models of 
the first type, e.g. ISO 25010, describe and structure general concepts that constitute 
high quality software. Most of them, however, lack 
the ability to be used for actual quality assessment or improvement.
The second kind of quality models is tailored for
specific domains, certain architectural paradigms (e.g.\ SOA) or single aspects
of software quality (e.g.\ reusability). 
They allow concrete assessments but often miss the
connection to higher level quality goals. Thus they make it difficult
to explain the importance of quality problems to developers or sponsors and
to quantify the economic potential of quality improvements. 
Because these specific
models usually do not cover the full spectrum of software quality, 
they impede the proliferation of a common understanding of software 
quality in the software industry. 

A similar gap also exists for quality assessment methods. 
Effective quality management requires not only a definition of quality 
and the measurement of individual properties but also
a method for assessing the overall quality of a software product 
based on the measured properties. 
Existing quality models either miss such 
assessment support completely or provide procedures that are 
too abstract to be operational (e.g.\ ISO~25040)
or not based on a solid theoretical basis (e.g.\ \cite{Bansiya2002}).
In consequence, quality assessment is inhibited and likely to produce 
inconsistent and misleading results; in particular if the
required assessment expertise is missing.

\subsection{Research Objective}

Our aim is a quality model for software that is both widely applicable 
and highly operationalised to
provide the missing connections between
generic descriptions of software quality attributes and specific
software analysis approaches.
The required operationalisation implies the integration of existing domain- 
and language-specific tools, manual analyses
and a soundly defined assessment method.

To achieve this goal, software quality experts from both academia 
and industry in Germany joined forces within the Quamoco
research project. The project consortium consists of Technische 
Universit\"at M\"unchen, SAP, Siemens, Capgemini, Fraunhofer 
IESE and itestra. In total, these partners spent 558 person months
on the project.

\subsection{Contribution}

Our work provides four major contributions:
First, we developed
a meta model for software quality, which covers the full 
spectrum from structuring quality-related concepts to defining
operational means to assess their fulfilment in a specific
environment. 
Second, for the actual quality assessment we contribute
a clearly defined assessment method that integrates with the
meta model. Third, the base quality model instantiates the meta model and captures
knowledge on how to conduct a basic quality assessment
for different kinds of software. 
At present, we have elaborated the base model  in depth for the languages Java and C\#. 
Fourth, we performed an initial validation of the model
with real software systems, which showed the correspondence
of the assessment results with expert opinions.

\section{Related Work}
\label{sec:relatedwork}

Quality models have been a research topic for several decades and a large number of 
quality models has been proposed~\cite{2009_klaes_QM_landscape}. The first work dates 
back to the late 1970s, when Boehm~et~al.\ described quality characteristics and their 
decomposition~\cite{1978_boehmb_software_quality}. The 1980s saw the first need for 
custom quality models and the first tool support. Until then, the quality models simply 
decomposed the concept \emph{quality} in more tangible quality attributes. In the 
1990s more elaborate ways of decomposing quality attributes were introduced by 
distinguishing between \emph{product components}, which exhibit \emph{quality carrying 
properties} and externally visible \emph{quality attributes}~\cite{Dromey.1995}. Later, 
Kitchenham~et~al.~\cite{Kitchenham.1997} acknowledged the need for an explicit meta model 
for quality models.

Based on the early quality models, the standard ISO~9126 was defined in 1991. As shown 
by various critiques (e.g.~\cite{deissenb:icsm07,AlKilidar.2005}) the used decomposition 
principles for quality attributes are often ambiguous. Furthermore, the resulting quality 
attributes are not specific enough to be directly measurable. Although the recently 
published successor ISO~25010 has several improvements, the overall critique is still valid. 
Our survey~\cite{2009_wagners_quality_models_practice,Wagner2010} shows that 
less than 28\% of the companies use these standard models and 71\% of them have 
developed their own QMs. Another weakness of these quality 
models is that they do not specify how the quality attributes should be measured 
and how measurement results can be aggregated to achieve an overall quality assessment for 
a system.

Although not embedded in an operationalised quality model, a large number of
tools for quality analysis are available: bug pattern identification (e.g.\ FindBugs, 
Gendarme, PC-Lint), coding convention checkers (e.g.\ Checkstyle), clone detection 
tools and architecture dependency/cycle analysis tools. These tools focus on specific 
aspects of software quality and fail to provide comprehensive 
quality assessments. Moreover, they are not explicitly and systematically linked to a quality model.

One can use the measurement data generated by these tools as input for dashboard tools 
(e.g.\ QALab, Sonar and XRadar). Their goal is to present an overview of the 
quality data of a software system. Nevertheless, they also lack an explicit connection between 
the metrics used and the required quality attributes. This results in a missing 
explanation of the impacts of found defects on software quality and in missing rationales for 
the used metrics.

A comprehensive approach is taken by the research project Squale~\cite{MordalManet.2009}. They develop an explicit quality model describing 
a hierarchical decomposition of the ISO~9126 quality attributes. The model contains formulas 
to aggregate and normalise metric values. Regarding the quality model, the main difference 
to our approach is that the model of Quamoco uses a product model to structure the quality 
factors. Based on the quality model, they provide tool support for evaluating software products. 
The measures and the quality model are fixed within these tools. In contrast, Quamoco 
offers an editor to create and manage quality models and the Quamoco tool chain allows for 
a flexible configuration and integration of measurement tools and even manually collected 
data.

In our prior work, we have investigated different ways of describing quality and
classifying metrics, e.g.\ activity-based quality models~\cite{deissenb:icsm07} and 
technical issue classifications~\cite{Plosch.2009}. Based on this work, we developed 
a meta model for quality models and evaluated it regarding expressiveness~\cite{Klas.2010}.
Regarding quality assessments and tool support for it, we experimented with different 
approaches~\cite{Plosch.2008,wagner:ist10,Lochmann.2011}. Based on the gained 
experience, we developed our tool-support~\cite{deissenboeck2011quamoco}. This 
paper describes the complete approach and an empirical validation.

\section{Quality Model Concepts}
\label{sec:concepts_meta}

The first challenge to address the gap between abstract quality
attributes and concrete assessments is to formalise general concepts for 
quality models by a suitable meta quality model. 
After we
describe how we use quality models, we explain each of the concepts briefly and reference 
which problems they solve. Finally, we
combine the concepts into a meta model to show the complete picture. These concepts and
the meta model have been developed in three iterations over three years with corresponding
evaluations~\cite{Klas.2010}.

\subsection{Usage of the Quality Models}

Most commonly, we find quality models reduced to just reference taxonomies or implicitly
implemented in tools. As explicit and living artefacts, however,
they can capture general knowledge about software quality, accumulate knowledge from
applying them in projects and allow to define a common understanding of quality in a specific 
context~\cite{2004_marinescur_oo_analysis,deissenb:icsm07,heitlager07,luckey10}.

We aim to use this knowledge as basis for quality control. 
In the control loop, the quality model is the central element for identifying quality goals, assessing these
goals, analysing defects and reworking the software product based on the analysis results. 
The quality model is useful to define what we need to measure and how we can interpret it to 
understand the state of quality of a specific product.
A single source of quality information avoids redundancies and inconsistencies in diverse quality 
specifications and guidelines.

On top of that, the model
itself helps us to establish suitable and concrete quality requirements. The
quality model contains quality knowledge that we need to tailor for the
product to be developed. This includes removing unneeded qualities as well as adding new
or specific qualities.

\subsection{General Concepts}

The previous work of all Quamoco partners on quality models, our joint discussions and 
experiences with earlier versions of the meta model brought us back to the basic concept
of a \emph{factor}. A factor expresses a \emph{property} of an \emph{entity}, which is similar to what 
Dromey~\cite{Dromey.1995} called \emph{quality carrying properties} of \emph{product components}. 
We describe with \emph{entities} the things that are important for quality and
with \emph{properties} the attributes of the things we are interested in. Because this concept of a
factor is rather general, we can use it on different levels of abstraction. We have concrete factors such as the cohesion of classes as well as abstract factors such as the portability of the product. 

To clearly describe quality from an abstract level down to concrete measurements, we explicitly
distinguish between the two factor types \emph{quality aspects} and \emph{product factors}. 
Both can be refined to sub-aspects and sub-factors.
The quality aspects express abstract quality goals, for example, the quality attributes of
the ISO 9126 and ISO 25010, which always have the complete product as their entity. 
The product factors are measurable attributes of parts of the product.
We require that the leaf product factors are concrete enough, so we can measure
them. An example is the duplication of source code, which we measure with \emph{clone coverage}.
This clear separation helps us to bridge the gap between the abstract notions of quality and concrete
implementations.

Moreover, we are able to model several different hierarchies of quality aspects to
express different views on quality. Quality has so many different facets that a single quality
attribute hierarchy is not able to express it. Even in the recent ISO 25010, there are two quality
hierarchies: Product quality and quality in use. We can model both as quality aspect hierarchies.
Also other types of quality aspects are possible. We experimented with our own earlier work:
activity-based quality models~\cite{deissenb:icsm07} (similar to quality in use of ISO 25010)
and technical classifications~\cite{ploesch_et_al_2009}. We found that this gives us the flexibility to
build quality models tailored for different stakeholders.

To completely close the gap between abstract quality attributes and assessments, we need to 
set the two factor types into relation. The product factors have \emph{impacts} 
on quality aspects. This is similar to variation factors, which have impacts on quality factors in
GQM abstraction sheets~\cite{Solingen1999}. An impact is positive or negative and describes how
the degree of presence or absence of a product factor influences a quality aspect. This gives us a
complete chain from measured product factors to impacted quality aspects and vice versa.

We need product factors concrete enough to be measured so that
we can close the abstraction gap. Hence, we have the concept of measures for product factors.
A measure is a concrete description how a specific product factor should be quantified
for a specific context. For example, this can be counting the number of violations of the rule for Java 
that strings should not be
compared by ``==''. A factor can have more than one measure if we need separate
measures to cover the concept of the product factor. Moreover, we separate the measures
from their \emph{instruments}. The instruments describe a concrete implementation of a measure. For the
example of the string comparison, an instrument is the corresponding rule as implemented in the
static analysis tool \emph{FindBugs}. This gives us additional flexibility to collect data for measures
manually or with different tools in different contexts. Overall, the concept of a measure also
contributes to closing the gap between abstract qualities and concrete software as it is possible to
trace down from the quality aspects over product factors to measures and instruments.

With all these relationships with measures and instruments, it is possible to 
assign evaluations to factors so that we can aggregate from measurement results (provided by instruments) to a complete quality assessment. There are different possibilities to implement that. We will describe a quality assessment method using these concepts later in Section~\ref{sec:quality_assessment}. Moreover, we
can go the other way round. We can pick quality aspects, for example, ISO 25010 quality attributes,
which we consider important and costly for a specific software system and trace down to what product factors
affect it and what are measures for that (cf.~\cite{wagner:ist10}). This way, we can
concentrate on the product factors with the largest impact on these quality aspects. It gives us
also the basis for specifying quality requirements, for which we developed an explicit
quality requirements method~\cite{Ploesch_et_al_2010,Lochmann2010}.

Building quality models in such detail results in large models with hundreds of model elements.
Not all elements are important in each context and it is impractical to build a single quality model
that contains all measures for all relevant technologies. Therefore, we introduced a modularisation
concept, which allows us to split the quality model into \emph{modules}. For that we have the \emph{root} 
module, which contains general quality aspect hierarchies as well as basic product factors and
measures. In additional modules, we extend the root module for specific technologies, such as
object-orientation, programming languages, such as C\#, and domains, such as embedded systems.

The modules enable us to choose appropriate modules and extend the quality model
by additional modules for a given context. To adapt the quality model for a specific company or
project, however, this is still too coarse grained. Hence, we also developed an explicit adaptation
method, which guides a quality manager in choosing relevant quality aspects, product
factors and measures for the current project~\cite{2011_klaes_QM_adaptation}.

\subsection{Meta Model}

\begin{figure}[htb]
\includegraphics[width=\columnwidth]{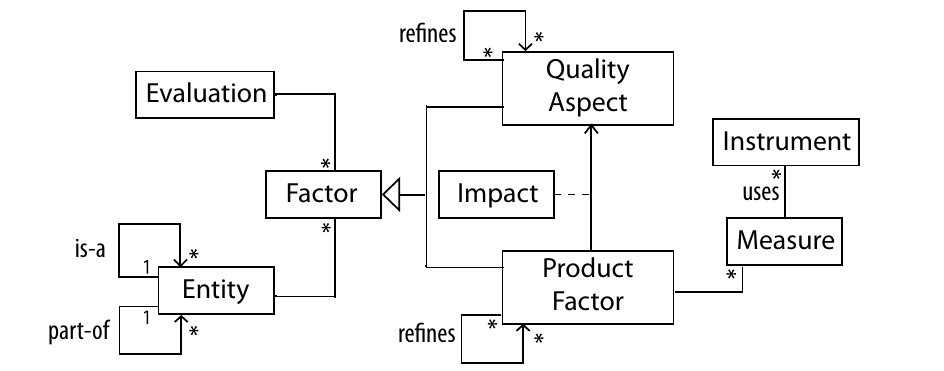}
\caption{The meta quality model (excerpt)}
\label{fig:metamodel}
\end{figure}

To precisely specify the general concepts described so far, we modelled them in a
meta model. The core elements of the meta model are depicted as an (abstracted) UML class diagram in Figure~\ref{fig:metamodel}.
Please note that we left out a lot of details such as the IDs, names and descriptions of each
element to make it more comprehensible. At the centre of the meta model resides the \emph{Factor} with
its specialisations \emph{Quality Aspect} and \emph{Product Factor}. Both can be refined and, hence, produce
separate directed acyclic graphs. An \emph{Impact} can only go from a \emph{Product Factor} to a \emph{Quality Aspect}.
This represents our main relationship between factors and hence allows us to specify the core quality concepts.

The \emph{Factor} always has an associated \emph{Entity}, which can be in a is-a as well as a
part-of hierarchy. For example, in an object-oriented language, a method is part of a class and is a
kind of source code. The property the \emph{Factor} describes of an \emph{Entity} is expressed
in the \emph{Factor}'s name. Each factor has also an associated \emph{Evaluation}. It specifies how
to evaluate or assess the \emph{Factor}. For that we can use the evaluation results from sub-factors
or -- in the case of a \emph{Product Factor} -- the values of associated \emph{Measure}s. 
A \emph{Measure} can be associated to more than one \emph{Product Factor} and has potentially several
instruments that allow us to collect a value for the measure in different contexts, e.g.\ with a
manual inspection or a static analysis tool.

We modelled this meta model with all details as an EMF\footnote{Eclipse Modeling Framework,
\url{http://emf.eclipse.org/}} model, which then became the basis for the quality model editor
 (see Section~\ref{sec:tool_support}).

\section{Base Model}
\label{sec:base_model}

Our main objective for the base model is to describe software quality in a way 
that allows tool-supported quality assessment and is applicable to a 
wide range of software products. 
To reach this goal, 
software quality experts from both academia and industry conducted a 
series of workshops over three years to collaboratively transfer their 
knowledge and experience into the structure described in section 
\ref{sec:concepts_meta}. The workshops 
covered the whole spectrum from full consortial meetings
to small, specialised teams creating and extending single modules.
The resulting QM represents our consolidated view on
the quality of software source code and is generally applicable to any
kind of software. By providing in-depth modelling, including particular
analysis tools as instruments for the assessment of Java and C\# systems,
it allows for 
comprehensive, tool-supported quality assessment without requiring large 
adaptation or configuration effort. Because this model constitutes the 
basis for further specialisation and adaptation, we call it the \emph{base model}.

\subsection{Contents}

The Quamoco base model is a comprehensive selection of factors and measures 
relevant for software quality assessment. In total, it comprises 112 entities 
and 286 factors. Since some factors are used for structuring purposes rather 
than quality assessment, only 221 factors have evaluations assigned. 
Of these, 202 factors define impacts on other factors, leading to a
total of 492 impacts. 
Since the model provides operationalisation for different programming 
languages (cf. Section~\ref{sec:base_model-modularization}), it contains considerably 
more measures than factors: In total, there are 194 measured factors and
526 measures in the model. 
For these measures, the model contains 542 instruments, 
which split up into 8 manual ones and 536 that are provided by one of 12 
different tools. The tools most relied on are FindBugs (Java, 361 rules 
modelled) and Gendarme (C\#, 146 rules). Other tools integrated into our model 
include PMD (Java, 4 rules) and several clone 
detection, size, and comment analyses that are part of the quality 
assessment framework.

In the following, we present example factors including their
respective measures and impacts to illustrate the structure
of the base model.

\subsubsection{Rules of Static Code Analysis Tools}

As described above, the largest fraction of measures refers to static
code analysis tools. One example is the FindBugs rule 
FE\_TEST\_IF\_EQUAL\_TO\_NOT\_A\_NUMBER, which scans Java code
for equality checks of floating point values with the \emph{Double.NaN} constant.
The Java language semantics defines that nothing ever equals to \emph{NaN},
not even \emph{NaN} itself, so that \emph{(x == Double.NaN)} is always false. 
To check whether a value is not a number, the programmer has to call \emph{Double.isNaN(x)}.
This rule is an instrument to the \emph{Doomed test for equality to NaN}
measure, which measures the factor \emph{General Expression Applicability}
for comparison expressions, along with a couple of other measures.
This factor in turn influences both \emph{Functional Correctness} and
\emph{Analysability}.

\subsubsection{Established Generic Factors}

Rule-based code analysis tools cannot detect every kind of quality problems.
Therefore, the base model also contains product factors based on 
established research results, metrics and best-practices. Identifiers have
been found to be essential for the understandability of source
code. Whether identifiers are used in a concise and consistent
manner can only partly be assessed automatically~\cite{2006_deissenboeckf_naming}.
Therefore, the factor \emph{Conformity to Naming Convention} for source
code identifiers contains both automatic checks performed by
several tools and manual instruments to assess whether identifiers
are used in a consistent and meaningful way.
Another established factor related to software quality is code cloning.
Source code that contains large amounts of clones was shown to be
hard to understand and maintain~\cite{Juergens.2009}.
The concept of code cloning is represented in the factor 
\emph{Duplication} of \emph{Source Code}, which has negative impacts
on both analysability and modifiability. It is measured by
\emph{clone coverage} as well as \emph{cloning overhead}.

\subsection{Modularisation} 
\label{sec:base_model-modularization}
According to the modularisation concept introduced in Section~\ref{sec:concepts_meta},
the base model is structured into several modules. 
In the base model, a module \emph{root} contains the definitions 
of quality aspects. For each programming language in the quality model an own module 
was introduced. For object-oriented programming languages (Java, C\#) an intermediate
module \emph{object-oriented} defines usual concepts of object-oriented programming 
languages such as classes or inheritance.

We used the modularisation concept to integrate individual analysis tools for
the programming languages. In the module \emph{object-oriented}, we defined a large number of
general metrics without connections to concrete tools (e.g.\ number of classes). 
The module for Java defines a tool for measuring the number of classes in Java systems. This 
way, we support a separation between general known concepts and specific instruments.

The explicit definition of modules provides several benefits to us: First, it enables
us to separately and independently work on modules for different technologies and domains.
Second, it allows us to explicitly model the commonalities and differences between several 
programming languages. This is visible in the reuse of factors of the module \emph{object-oriented} 
in the modules \emph{Java} and \emph{C\#}: The common module defines 64 factors, \emph{Java} adds 
only 1 and \emph{C\#} only~8 language-specific factors. 

We also used the modularisation concept to connect domain-specific quality models
with the root model. The industry partners defined their own models for their domain.
For example, itestra defined a model for information systems. It extends the root model
and adds factors describing quality characteristics of database schemas and tables.

\section{Quality Assessment Approach}
\label{sec:quality_assessment}

A QM specifies quality in terms of relevant properties of software artefacts and associated measures. Yet, to support assessing product quality the QM needs to be associated with an approach to synthesise and interpret the measurement data collected for the product. In this section, we specify a quality assessment method applicable for Quamoco QMs.

\subsection{Practical Challenges}
In practice, there are a number of specific challenges that quality assessment must address in addition 
to the challenges of quality modelling. To identify these challenges and to determine how existing quality 
assessment methods address them, we performed a systematic literature review and a survey among 
members of the Quamoco consortium. Details on the design, execution and results of that survey are 
beyond the scope of this paper and will be published separately.

Among the most important 
requirements were that a software quality assessment should be comprehensible to software decision makers, 
combines quality preferences of different groups of stakeholders, copes with incomplete information 
and allows for mutual compensation between multiple (potentially contradictory) quality 
aspects. We observed that the structure of the quality assessment problem corresponds to the problems 
addressed by Multicriterial Decision Analysis (MCDA) \cite{Keeney1993}. As a result of a literature review 
we decided to adapt principles of Multiple-Attribute Utility/Value Theory (MAUT/MAVT) \cite{Keeney1993}, as it meets 
most of the requirements and can be easily adjusted to meet all the requirements.

\subsection{Quality Assessment Method}

The Quamoco quality assessment method models the preferences of decision makers for a product's quality 
using the concept of utility. Utility quantifies the relative satisfaction of a decision maker concerning the quality 
of a software product characterised by specific measurable factors. 

While measures provide objective values without preferences, the utility defines the quality preferences of a decision maker in that for any two products the one with higher utility is preferred. For example, for the purpose of assessing software maintainability we may consider the factor \emph{density of comments in the software source code}, measured as the percentage of comment lines in the complete source code. The utility of this factor would not be monotonic. From the perspective of maintaining software code, we would prefer more code comments only up to a certain threshold (utility increases with increasing comment density) and drop after exceeding this threshold (utility decreases with increasing comment density). For each measurable factor, we can define a different mapping between the factor measurements and corresponding utilities using a utility function. In case of multiple factors, the total product utility is a synthesis of the utilities of all individual factors. Within the hierarchical structure of the base model, the utility function defines a mapping between the factor's measurement values and its utility. At higher levels of the QM hierarchy, the utility of a factor results from a synthesis of the utilities assigned to all its direct sub-factors.

\begin{figure}[htbp]
\begin{center}
\includegraphics[width=\columnwidth]{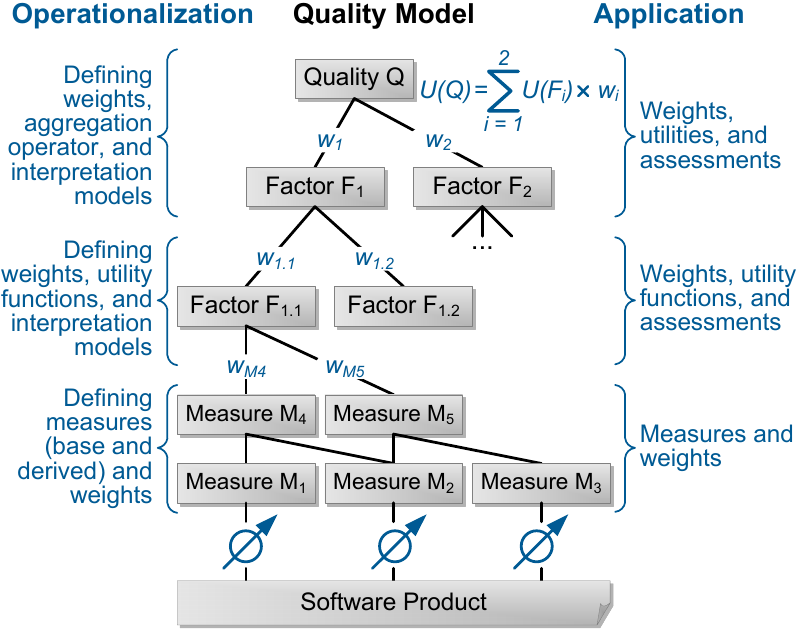} 
\caption{Overview of the quality assessment approach}
\label{fig:ass1}
\end{center}
\end{figure}

Figure~\ref{fig:ass1} shows the quality assessment activities within the hierarchical structure of the base model. On the left side of the figure, there are key activities and outputs needed for operationalising the model to perform assessments. These activities and their outputs correspond to the generic process of MCDA (e.g.\ \cite{Dodgson2000}). Analogically, on the right side of the figure, there are the corresponding activities performed during the assessment of a specific product and their outputs. Subsequently, we discuss the objective and concrete procedure applied for each activity couple when we operationalise and apply the base model for quality assessments.

\subsubsection{Defining Measures / Measurement} 
During the operationalisation, we have to associate measures for all input data needed for the quality assessment with appropriate measurement instruments. Each measure was reviewed by two measurement experts to decide on an appropriate normalisation measure based on a defined set of rules. For instance, base measure \textit{$M_1$: Doomed test for equality to NaN} is normalised by base measure \textit{$M_2$: Lines of code} into derived measure $M_4$, through which measurements of $M_1$ become comparable between software systems of different sizes. The measures quantify the lowest-level factors in the QM hierarchy. During the application, objective measurement data are collected. For example, for the source code of the Java Platform, version 6, we obtain $M_1 = 6$, $M_2 = 2,759,369$, and consequently $M_4 = 2.17 \cdot 10^{-6}$. To cope with incomplete measurement data, the approach uses interval arithmetic to determine the range of possible outcomes for factors.

\subsubsection{Defining Utility Functions / Scoring} 
During the operationalisation, we defined a utility function for each measure of a factor at the lowest level of the hierarchical QM. These functions define the utility each measure has in the context of the factor it is associated to. The factor's utility is then defined as the weighted sum of the utilities of all measures connected to the factor. To assure the understandability of the evaluation, we used only simple linear increasing and decreasing functions with two thresholds \emph{min} and \emph{max} that determine when the factor is associated with the minimal (0) and maximal utility (1). After experts decided on the type of function (decreasing or increasing), we determine the thresholds for the function using Equation 1 on the normalised measurement values for a set of at least 10 baseline systems. For the Java specific-part, we used, for instance, data from a sample of 110 open source software systems.

\begin{equation}
\begin{array}{l}\displaystyle
	\mbox{IF } \left| \left\{ s_{i=1...n} : s_i > 0 \right\} \right| < 5 \mbox{ THEN} \\
	\quad \mathit{min} = 0, \mathit{max} = 0.00000001 \\
	\mbox{ELSE}\\
	\quad \begin{array}{ll}\mathit{max} = \mathit{max}&\left(\left\{ 
	s_i : s_i \leq 
	\begin{array}{l}
	 Q3(\{s_i : s_i \neq 0\})
\\+ 1.5 \cdot \mathit{IQR}(\{s_i\}) 
\end{array}
\right\}\right), \\
\mathit{min} = \mathit{min}&\left(\left\{ 
	s_i : s_i \geq 
	\begin{array}{l}
	 Q1(\{s_i : s_i \neq 0\})
\\- 1.5 \cdot \mathit{IQR}(\{s_i\}) 
\end{array}
\right\}\right) \\
\end{array}\\
\mbox{END}
\end{array}
\end{equation}

where $s_i = S(F_x)$ for baseline system $i$ and where $Q1$ and $Q3$ represent the 25\% and 75\% percentiles.
$\mathit{IQR} = Q3 - Q1$ represents the inter quartile range.

The equation assures that for measures with a limited number of data points different from zero a simple jump function at 0 is defined. Else, the minimum and maximum non-outlier values are used as thresholds. For example, we obtained for $M_4$: $\mathit{min} = 0$, $\mathit{max} = 8.5 \cdot 10^{-6}$. Two measurement experts reviewed the automatically determined thresholds for each measure together with supporting descriptive statistics for plausibility.
During the application, we calculated the defined evaluation function on the measurement data of the assessed system. This operation involves evaluating the utility of a factor at the lowest level of the QM. For the Java Platform, version 6, we obtain, for instance, $U(M_4) = \mathit{Eval}(M_4) = 0.74$

\subsubsection{Defining Factor Weights \& Aggregation Operator / Aggregation} During operationalisation, the relative importance of adjacent elements of the QM has to be specified, where elements include factors and measures directly associated to lowest-level factors. We extracted the relative importance of quality aspects from the results of our survey\cite{Wagner2010}.
For other relationships, mixed teams consisting of industrial and academic experts determined the relative importance. Weights for adjacent elements must be between 0 and 1 and sum up to 1. To support the efficient definition of weights, we used the Rank-Order Centroid method \cite{Barron1996} to calculate the weights of each factor automatically based on a relevance ranking between sibling factors provided by the teams using the \textit{Swing} approach \cite{Edwards1994}. 

In our case, $M_4$ was rated as less important for $F_{1.1}$ (General expression applicability of comparison expressions) than the second measure $M_5$: Floating point equality and, therefore, obtained the weight $w_{M_4} = 0.25$. 
During the model application, we use the weights within the bottom-up synthesis of factor utilities along the hierarchy of the QM. For this purpose, we define an appropriate aggregation operator. We use a weighted sum operator as an easily understandable and relatively reliable aggregation approach. 
For instance, we obtain for $U(F_{1.1}) = w_{M_4} \cdot U(M_4) + w_{M_5} \cdot U(M_5) = 0.25 \cdot 0.74 + 0.75 \cdot 0.89 = 0.85$ and for $F_1$ (functional correctness) $U(F_1) = w_{1.1} \cdot U(F_{1.1}) + w_{1.2} \cdot U(F_{1.2}) = 0.02 \cdot 0.85 + ... = 0.82$.

\subsubsection{Defining / Applying Interpretation Models}
These activities support the decision maker in interpreting the factor's utility, for example if it is good or bad. The objective of interpretation is to map the ratio-scale utility, for instance, onto a more intuitive, ordinal scale such as school grades or traffic lights.


\section{Tool Support}
\label{sec:tool_support}
The base model was designed in a way
that it can be used as provided without any modifications for any software
project, regardless of the application domain. Thus, the model as well as the 
assessment method is ready to use and can be applied with minimal effort on a project.

Additionally, the Quamoco project contributes an integrated tool chain for both
quality modelling and assessment \cite{deissenboeck2011quamoco},
available from the Quamoco
website\footnote{\url{https://quamoco.in.tum.de/wordpress/?page_id=767&lang=en}}. The
tooling consists of the quality model editor and the quality
assessment engine which we describe in the following.

\subsection{Quality Model Editor} 

The quality model editor is built on the Eclipse Platform and the
Eclipse Modeling Framework. It allows modellers to edit QMs
conforming to the Quamoco meta quality model. According to the
modularisation concept (cf. Section
\ref{sec:base_model-modularization}), each module of a QM is stored in
a separate file. This enables concurrent work on a QM by multiple
users. The content of the model can be navigated by different tree
views that allow form-based editing of the attributes of model
elements.

Validation during editing helps the modeller to create models adhering
to meta model constraints, consistency rules and modelling best
practices. A simple validation rule checks for unreferenced model
elements. A more sophisticated rule ensures that for model elements
referenced in other quality modules an appropriate \emph{requires}
dependency between the modules is defined. The editor employs the
Eclipse marker mechanism for displaying error and warning messages in
a list and provides navigation to affected elements. The user is
further assisted by an online help feature that displays
context-sensitive help content depending on the current selection in
the editor. The help texts explain the concepts of the meta quality
model and contain a guideline with best practices for quality
modelling.

\subsection{Quality Assessment Engine} 

The quality assessment engine is built on top of the quality
assessment toolkit ConQAT\footnote{\url{http://www.conqat.org/}} which
allows to create quality dashboards integrating diverse quality
metrics and state-of-the-art static code analysis tools.

The connection between the quality modelling and the assessment is
achieved by an automated generation of a ConQAT analysis configuration
from a QM. For the assessment of a software system, the quality
assessment engine is provided with the QM, the code of the software
system to assess, the generated ConQAT configuration and, optionally,
manual assessment results stored in an Excel file. This allows the assessor to
extend the tooling with custom analyses needed for the evaluation of
an own QM.

The assessment result can be inspected within the editor in the
hierarchy of the QM. Moreover, a treemap visualisation of the results
allows to track down quality issues from abstract quality
characteristics to concrete measures. Finally, an HTML report allows
to inspect the results of an assessment from within a browser, thus
not requiring the tooling and the QM. The quality assessment engine
can also run in batch mode, which enables the integration in a
continuous integration environment. Thereby, a decay in quality can be
detected early.


\section{Empirical Validation}
\label{sec:validation}

We validate the quality assessments grounded on the base model using two
research questions:\\
\textbf{RQ 1}: Can the base model be used to detect quality differences between different systems or subsystems?

\noindent\textbf{RQ 2}: Can the base model be used to detect quality improvements over time in a software system?

\subsection{Comparison of Software Products and Subsystems}

To answer RQ~1, we have to evaluate whether the base model provides valid assessment results, meaning that the assessment results are in concordance with the results obtained by another independent and valid approach for assessing product quality. We check RQ~1 for (a) products and (b) subsystems, because in practice these are the two most important applications for quality assessments:  compare products with respect to quality and identify parts of a system that need further quality improvement.

\paragraph{Design} To evaluate the validity of the quality assessments, we need an independently obtained criterion for product quality that we can compare with our assessment results. Since no measurement data were available that directly measure the quality or the quality aspects of interest for a set of products or subsystems that we can assess using the base model, we utilised as the independent criterion expert-based quality judgments. (a) For the comparison of different software products, we used the rating provided in the \emph{Linzer Software-Verkostung} \cite{Gruber2010} for a set of five open source products. The rating provided is a ranking of the five systems based on a combination of ratings provided independently by nine experienced Java experts. (b) For the comparison of different subsystems of one product, we used five subsystems of a business software system developed by one of the industry partners and a ranking provided by an expert from the company familiar with the five assessed subsystems. 

To measure validity and ensure comparability with other studies, we make use of the validity criteria proposed in the IEEE standard 1061 for validating software quality metrics. The standard proposes a set of criteria but most of them assume that the collected measures and the independent criterion both use an interval or ratio scale. In our case, the results of the base model assessments are provided as a value characterising the product quality between 1 (best possible) and 6 (worst possible) and the assessment results of the expert judgements provided on an ordinal scale as a ranking from best (1) to worst (5) product or subsystem, respectively. Consequently, we had to limit our investigation to the validity criterion \emph{consistency} (cf. IEEE 1061), which can be applied on ordinal scale data. It characterises in our case the concordance between a product ranking based on the assessments provided by our model and the ranking provided independently by (a) a group of experts / (b) an expert. This means that we determine whether the base model can accurately rank the set of assessed products / subsystems with respect to their quality (as perceived by experts).

Following the suggestion of IEEE 1061, we measure consistency by computing the Spearman's rank correlation coefficient (r) between both rankings, where a high positive correlation means high consistency between the two rankings. Since we want to check whether a potentially observed positive correlation is just due to chance or is a result of using an appropriate quality model, we state the hypotheses $H_{1_A}$ and $H_{2_A}$ (and corresponding null hypotheses $H_{1_0}$ and $H_{2_0}$). We test both with the confidence level 0.95 ($\alpha$ = 0.05):

$H_{1_A}$: There is a positive correlation between the ranking of the systems provided by the base model (BM) and the ranking of the systems provided by the experts during the Linzer Software-Verkostung (LSV).

$H_{2_A}$: There is a positive correlation between the ranking of the subsystems provided by the base model (BM) and the ranking of the subsystems provided by the company expert (Exp).

\begin{eqnarray*}
H_{1_A}: r (\mbox{rankingBM}, \mbox{rankingLSV} ) > 0 \\   
\mathrm{i.e.}, H_{1_0}: r (\mbox{rankingBM}, \mbox{rankingLSV}) \le 0 \\
H_{2_A}: r (\mbox{rankingBM}, \mbox{rankingExp} ) > 0 \\   
\mathrm{i.e.}, H_{2_0}: r (\mbox{rankingBM}, \mbox{rankingExp}) \le 0 
\end{eqnarray*}

\paragraph{Execution} (a) During the study, we used the base model to assess the quality of five open source products for which independent expert-based assessment results of the Linzer Software-Verkostung were available: JabRef, TV-Browser, RSSOwl, Log4j and Checkstyle. We ordered the assessed products by the results for their overall quality provided by the base model and compared them with the ranking provided by the Linzer Software-Verkostung. (b) Moreover, we employed the base model to assess the overall quality and maintainability of the five selected subsystems of a business software developed by one of the industry partners. Based on these values, we ranked the subsystems with respect to maintainability and the overall quality. Independent of this, we interviewed an expert in the company and asked him to rank the subsystems with respect to maintainability and overall quality. When he could not decide on the order of two subsystems, he provided the same rank for the two subsystems. In both studies, we followed a cross-validation approach (i.e., none of the assessed systems / subsystems were part of the set of systems used to calibrate the base model). 

\paragraph{Results} Table \ref{tab:verkostung} shows the assessment results using the base model and the resulting product ranking as well as the ranking of the Linzer Software-Verkostung. The calculated Spearman's rho correlation is r = 0.975, which is close to a perfect correlation of 1. Hypothesis $H_{1_A}$ can also be accepted on a high level of significance (p=0.002) meaning that there is a significant positive correlation between the ranking provided by the base model and the ranking provided by the Linzer Software-Verkostung. 

\begin{table}[htp]
\caption{Comparison of the assessment results and the results of ``Linzer Software-Verkostung''}
\begin{center}
\begin{tabular}{lrrrr}
\hline
Product & LOC & Result BM &	Rank BM & Rank LSV\\
\hline
Checkstyle &	 57,213 & 1 (1.87)	& 1 &	1\\
RSSOwl &		 82,258 & 3 (3.14)	& 2 &	3\\
Log4j &			 30,676 & 3 (3.36)	& 2 &	2\\
TV-Browser &	125,874 & 4 (4.02)	& 4 &	4\\
JabRef &		 96,749 & 5 (5.47)	& 5 &	5\\
\hline
\end{tabular}
\end{center}
\label{tab:verkostung}
\end{table}%

The assessment results for the subsystems are shown in Table~\ref{tab:fivesubsystems}. The calculated Spearman's rho correlation for the overall quality is r = 0.32, which is a positive but only moderate correlation. Hypothesis $H_{2_A}$ cannot be accepted (p=0.30) meaning that we cannot show the statistical significance of the observed positive correlation. The agreement between the expert and the base model results is higher (r = 0.67) when considering maintainability, but also not statistical significant (p=0.11).

\begin{table}[htp]
\caption{Comparison of the assessment results of five subsystems and an expert's opinion}
\begin{center}
\begin{tabular}{lrrrr}
\hline
  					 & Rank Exp & Rank Exp & Rank BM & Rank BM \\
Subsystem 	 & Quality & Maint. & Quality & Maint. \\
\hline
Subsystem A   & 2 & 2 & 5 & 2 \\
Subsystem B   & 2 & 3 & 3 & 5 \\
Subsystem C   & 4 & 4 & 2 & 3 \\
Subsystem D   & 1 & 1 & 1 & 1 \\
Subsystem E   & 4 & 3 & 4 & 4 \\
\hline
\end{tabular}
\end{center}
\label{tab:fivesubsystems}
\end{table}%

\paragraph{Interpretation} (a) The assessments of the overall product quality for the five investigated systems turn out to be consistent and thus valid when compared to an independent criterion for quality, in this case provided in the form of an expert-based assessment. Although this conclusion is supported by a very high and statistically significant correlation, there are some threats to validity that need to be considered.

(b) For the investigated subsystems, the results are not that clear. The study hints that the Quamoco base model is useful to identify -- or at least narrow down -- the list of subsystems that need further quality improvement. However, since the correlation is only moderate, a larger sample of subsystems would be needed to draw statistical significant conclusions.

\paragraph{Threats to Validity} The most relevant threats we see are (1) we cannot guarantee that the criterion chosen for the validation, namely the expert-based quality rating, adequately represents the quality of the products/subsystems, this is especially relevant for the rating of the subsystems where only one expert rating was available. (2) The generalizability of our results is limited by the fact that the scope of the empirical validation is limited to five medium sized Open Source systems and five subsystems of one industrial software system written in Java.

\subsection{Making Quality Improvements Visible}

In a second study we knew from the quality experts of the project that they had invested effort in enhancing the maintainability of the 
system. In this study from the automation domain (steel production) we analysed six versions of a Java software using our base model. The major goal was to validate, whether our base model would reflect the assumed quality improvements claimed by the quality managers (RQ~2). It is important to understand that the quality improvement actions were not based on any results of static code analysis tools but on the experience of the developers involved. Thus, as a side effect the validation should also show to some extent whether our base model reflects the common understanding of quality (in the context of maintenance) of experienced developers. 

\paragraph{Results} Table \ref{tab:quality_improvement} shows the assessment results using the base model and the corresponding calculated
quality grades. The quality experts stated that they explicitly improved the quality starting right after rollout of version 2.0.1.
There is no quality
related information available for versions 1.9.0 and 2.0.0.

\begin{table}[htp]
\caption{Quality improvements in an automation software project}
\begin{center}
\begin{tabular}{lr}
\hline
Version & Quality grade\\ 
\hline
1.9.0 & 4.15\\
2.0.0 & 3.34\\
2.0.1 & 3.63\\
2.0.2 & 3.42\\
2.1.0 & 3.27\\
2.2.1 & 3.17\\
\hline
\end{tabular}
\end{center}
\label{tab:quality_improvement}
\end{table}%

\paragraph{Interpretation} The results clearly show steady improvements (as expected) for the versions 2.0.2., 2.1.0 and 2.2.1. Our assessment model calculates a considerable improvement of 12.68\% from version 2.0.1 to version 2.2.1. This result reflects the expectations of the quality experts of the project. 

\paragraph{Threats to Validity} The most relevant threat we see is that we had only one industry project at hand where the quality experts
explicitly invested in quality without using static code analysis tools. The generalizability of this result is therefore limited.  

\section{Conclusions}
\label{sec:conclusion}

In practice, a gap exists between abstract quality definitions provided in common quality taxonomies, such as ISO 25010, and concrete quality assessment techniques and measurements \cite{2009_wagners_quality_models_practice}. Our overall aim is to close this gap by operationalised quality models. We have shown in this paper our four contributions to achieve this goal: (1) We developed an explicit meta model, which allows us to specify operationalised quality models by the flexible but well-defined concepts of \emph{factors}, \emph{impacts} between factors and \emph{measures} for assessing the factors. (2) Using this meta model, we built a broad, largely technology independent base model that we exemplarily operationalised for the programming languages Java and C\#. The freely available and extendable base model captures the most important product factors and their impacts to product quality as defined in ISO 25010. (3) We provided a quality assessment approach, which enables us to use the base model for transparent and repeatable quality assessments.
(4) We evaluated two aspects of the complete approach in empirical studies. We found that the assessment results fit to expert opinion but the strongest results are limited to the maintainability part of the model. In addition, we have developed extensive, open-source tool support for building operationalised quality models as well as performing the quality assessments.

By working on filling the gap in current quality models, we found several more directions of future work that should be followed. First, the base model and its evaluation concentrate on a small number of technologies so far. To be truly broad, we need to take further technologies and contents for the base model into account. Second, we are working on further empirical studies to understand the still existing weaknesses of our approach to further improve them. In particular, we work with all industry partners on \emph{drill-downs} where system experts rate the usefulness of the quality assessment results. 

\section*{Acknowledgment}

We are grateful to all present and former members of the Quamoco project team
as well as all the participants of our interviews and surveys. This work has been
supported by the German Federal Ministry of Education and Research under
grant number 01IS08023.

\small

\end{document}